\documentclass[
superscriptaddress,
nofootinbib,
bibnotes,
amsmath,amssymb,
aps, twocolumn,
pra,
]{revtex4-2}

\usepackage[utf8]{inputenc}
\usepackage{siunitx}
\usepackage[english]{babel}			
\usepackage{graphicx}				
\usepackage{amssymb,amsmath,mathtools, bbold}		
\usepackage{url}					
\usepackage{float}                  
\usepackage{caption}
\usepackage{braket}
\usepackage{bm}
\usepackage{hyperref}
\hypersetup{colorlinks=true,allcolors=blue}
\usepackage{lineno}

\DeclareCaptionLabelFormat{custom}{#1 \textbf{#2}}
\AtBeginDocument{\RenewCommandCopy\qty\SI}

\preprint{APS/123-QED}

\begin{document}

\makeatletter
\patchcmd{\frontmatter@RRAP@format}{(}{}{}{}
\patchcmd{\frontmatter@RRAP@format}{)}{}{}{}
\renewcommand\Dated@name{}
\makeatother
\date{Date: \today \\ \phantom{XXX} E-mail: Tommaso.Faleo@uibk.ac.at}

\title{Photonic hyperentanglement in polarisation and frequency via joint spectrum shaping}

\author{Tommaso Faleo}
\email{Tommaso.Faleo@uibk.ac.at}
\affiliation{Institut für Experimentalphysik, Universität Innsbruck, Technikerstr. 25, 6020 Innsbruck, Austria}
\author{Fabian Steinhauser}
\affiliation{Institut für Experimentalphysik, Universität Innsbruck, Technikerstr. 25, 6020 Innsbruck, Austria}
\author{Gregor Weihs}
\affiliation{Institut für Experimentalphysik, Universität Innsbruck, Technikerstr. 25, 6020 Innsbruck, Austria}
\author{Stefan Frick}
\affiliation{Institut für Experimentalphysik, Universität Innsbruck, Technikerstr. 25, 6020 Innsbruck, Austria}
\author{Robert Keil}
\affiliation{Institut für Experimentalphysik, Universität Innsbruck, Technikerstr. 25, 6020 Innsbruck, Austria}
\begin{abstract}

Hyperentanglement offers enhanced capacity for quantum information processing and communication protocols, especially in combination with robust high-dimensional degrees of freedom such as frequency-bin encoding. 
Here, we present a single-pass, unfiltered, down-conversion source of hyperentangled photon pairs in polarisation and frequency-bin degrees of freedom with dynamically tunable state dimension and composition at telecom wavelengths.
We achieve this by optimal tailoring of the photons' joint spectral amplitude via pump and nonlinearity shaping.
Using polarisation-resolved time-of-flight spectrometry and Hong-Ou-Mandel interference, we characterise the hyperentangled states and demonstrate for the polarisation component fidelities exceeding \qty{99}{\percent} averaged over frequency bins and concurrences above \qty{98}{\percent}. The degree of spectral entanglement, quantified by the Hong-Ou-Mandel visibility, is measured as \qty{90}{\percent}, well in line with numerical simulations.
This approach provides a scalable route toward high-dimensional quantum states for quantum communication and computing applications.

\end{abstract}

\maketitle

\section*{Introduction}

Photonic quantum technologies rely on the ability to generate, manipulate, and measure quantum states of light with high fidelity~\cite{O'Brien2009}.
Entangled photon pairs, in particular, serve as fundamental resources for quantum key distribution~\cite{Ekert1991, Kimble2008, Valencia2025}, quantum teleportation~\cite{Bouwmeester1997}, and quantum computing~\cite{Knill2001, Raussendorf2001, Bartolucci2023}.
While polarisation entanglement has been extensively studied and offers straightforward manipulation through standard optical elements, encoding in higher-dimensional degrees of freedom (DOF), such as spatial mode~\cite{Mair2001, Erhard2018, Lib2024}, time-bin~\cite{Brendel1999, DeRiedmatten2002, Yu2025}, pulse-mode~\cite{Ansari2018, Graffitti2020}, and frequency~\cite{Imany2018, Chang2025}, provides pathways to increased information capacity and noise resistance~\cite{Erhard2020}.

Hyperentanglement, wherein photon pairs exhibit entanglement in multiple DOF simultaneously, offers additional compelling advantages for quantum protocols~\cite{Kwiat1997, Barreiro2005, Deng2017}.
Hyperentangled states enable complete Bell-state analysis~\cite{Walborn2003, Schuck2006, Barbieri2007, Li2016, Gao2020, Zeng2020}, enhance the efficiency of quantum communication~\cite{Barreiro2008, Kim2021, Chapman2022, Zhong2024}, provide resources for entanglement purification and distribution over noisy channels~\cite{Simon2002, Sheng2010, Ecker2021, Huang2022, Ecker2022}, and give access to cluster states generation~\cite{Reimer2019}. 
In this context, hyperentanglement in the polarisation and frequency-bin DOF represents a particularly promising approach for quantum networking, combining the advantages of polarisation entanglement with the high dimensionality of frequency encoding and its natural compatibility with dense wavelength-division multiplexing infrastructure in fibre-optic networks~\cite{Kues2017, Siddarth2020, Alshowkan2021, Huang2022_qkd}.
In these schemes, entanglement is generated across discrete spectral modes, enabling the encoding of high-dimensional quantum states within the bandwidth of spontaneous parametric down-conversion (SPDC) or spontaneous four-wave mixing (SFWM) sources~\cite{Lukens2017}.
The discrete spectral modes are typically obtained by filtering the broadband spectrum of photon pairs or using microresonators~\cite{Lu2023_2,Lu2023}, which severely limits the heralding efficiency of the sources and, consequently, the success rate of the downstream protocols.
In contrast, engineering of the nonlinearity profile of quasi-phase-matched SPDC crystals has recently enabled the direct generation of frequency-bin-entangled states and polarisation-frequency hyperentangled states with pairwise anti-correlated frequency modes without the need for filtering~\cite{Morrison2022}.
Moreover, extending this approach to a two-dimensional shaping of the photon pairs' joint spectral amplitude (JSA) can scale up the number of spectral features and generate more complex frequency-bin states such as multi-mode and hybrid squeezed states~\cite{Drago2022}, two-qutrit states with both, on-diagonal and off-diagonal correlations~\cite{Shukhin2024}, and grid states with potential application in quantum error correction~\cite{Fabre2020}.

In this work, we experimentally demonstrate a source of hyperentangled photon pairs in the polarisation and frequency-bin DOF at telecom wavelengths (\qty{\approx 1550}{nm}), where the frequency modes of the pair are distributed and entangled in two dimensions, in both the diagonal and anti-diagonal directions, beyond the conventional correlation/anti-correlation dichotomy.
This is achieved by leveraging our recently developed scheme for flexible JSA engineering in SPDC sources~\cite{Faleo2025}.
Frequency-bin entanglement is implemented by combining two complementary strategies as theoretically proposed in Ref.~\cite{Drago2022}: programmable spectral shaping of a femtosecond pump laser to create a multi-Gaussian pump envelope function (PEF), and custom aperiodically-poled potassium titanyl phosphate (aKTP) crystals designed to produce a multi-Gaussian phase-matching function (PMF).
This combined approach enables direct, precise shaping of the frequency modes with high intra-bin spectral purity, without the need for a cavity or other filtering mechanisms, and with full control over the weights of the different modes.
Polarisation entanglement is simultaneously generated by placing the crystals within Sagnac loop interferometers.
We show the tunability and accuracy of our pump spectral shaping approach by producing three different (two- and four-mode) frequency-bin entangled states and demonstrate fidelities to maximally polarisation-entangled states \qty{\geq 99}{\percent} across all frequency bins, with concurrences \qty{\geq 98}{\percent}.
We characterise the hyperentangled states using polarisation-resolved time-of-flight spectrometry (TOFS), which enables joint measurement of spectral and polarisation correlations, and we verify the hyperentangled nature of the states by observing fine-grained and exchange-symmetry dependent fringes in Hong-Ou-Mandel (HOM) interference measurements~\cite{HOM1987}.

Section~\ref{sec:joint spectral amplitude shaping} introduces the concept of two-dimensional joint spectral shaping and our experimental setup for hyperentangled state generation.
We present the results of the spectral shaping, in the absence of polarisation-entanglement, in section~\ref{sec:spectral characterisation and hong-ou-mandel interference}, and we show the tunability of the approach in section~\ref{sec:spectral shaping tunability}.
The hyperentanglement characterisation obtained through polarisation-resolved TOFS and HOM interference is reported in sections~\ref{sec:polarisation entanglement in individual frequency bins} and \ref{sec:HOM hyperentangled state}, respectively. 

\begin{figure*}
    \centering
    \includegraphics{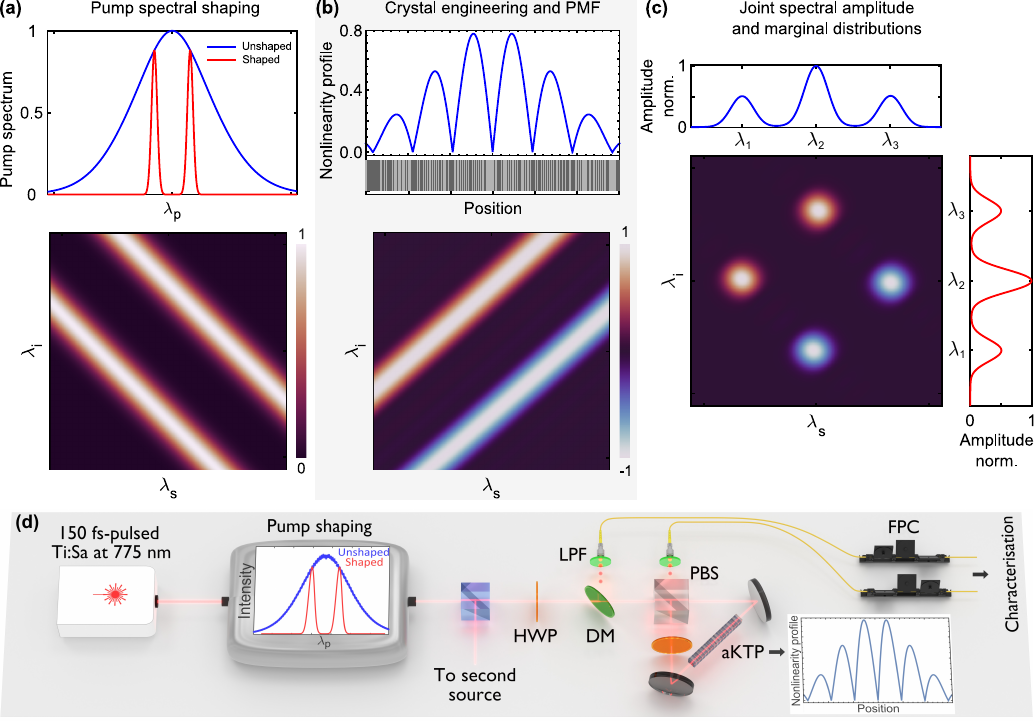}
    \caption{Joint spectral amplitude engineering and hyperentanglement generation setup. \textbf{(a)} Spectral shaping is employed to tailor a multi-Gaussian spectrum (red line) from the unshaped laser pump spectrum (blue line). The corresponding simulated multi-Gaussian PEF is shown on the bottom panel.
    \textbf{(b)} The nonlinearity profile (blue line) is shaped to achieve a multi-Gaussian PMF (see bottom panel) by engineering the crystal's poling pattern, as depicted in the inset. Light and dark grey regions have inverted nonlinearity signs. The specific pattern used here introduces a phase shift of $\pi$ between the antinodes of the PMF.
    \textbf{(c)} The resulting joint spectral amplitude shows fully separate frequency bins, as shown in the marginal distributions of $|\text{JSA}|$.
    Schmidt decomposition yields a Schmidt number $K=2.033$ due to the symmetry of the JSA.
    \textbf{(d)} A femtosecond (fs)-pulsed laser is spectrally shaped with a pulse shaper~\cite{Faleo2025} and the output is split at a beam splitter to pump two photon-pair sources, each consisting of a custom apodised-KTP crystal inside a Sagnac loop.
    A half-wave plate (HWP) controls the pump polarisation, determining whether separable or entangled photon pairs are generated.
    The hyperentangled down-converted photons are separated from the pump beam using a polarising beam splitter (PBS), a dichroic mirror (DM) and lowpass filters (LPF), and their polarisation state is tuned through fibre polarisation controllers (FPC).}
    \label{fig:fig1}
\end{figure*}

\section{\label{sec:joint spectral amplitude shaping}Joint Spectrum Shaping and Hyperentanglement Generation}

The JSA of photon pairs generated via SPDC is defined by the product of the PEF, determined by the pump spectrum, and the PMF, determined by the phase-matching conditions in the nonlinear crystal~\cite{Mandel1985, Grice1997}.
Engineering both functions enables precise control over the spectral correlations of the generated photon pairs.
In quasi-phase-matching conditions~\cite{Hum2007}, this can be achieved with custom design of the crystal's poling pattern to tailor the nonlinearity profile~\cite{Tambasco2016, Graffitti2018, Frick} combined with spectral shaping of the pump laser~\cite{Faleo2025}. 
Not only can this be used to obtain spectrally pure photon pairs, but it also allows the generation of more complex frequency-bin structures.
We will focus here on a $C_4$-rotational symmetric two-by-two frequency-bin structure, as shown in Fig.~\ref{fig:fig1}, but this approach can be extended to more frequency modes~\cite{Drago2022}, as shown in Sec.~\ref{sec:spectral shaping tunability}, limited by the accessible pump spectral bandwidth.

Figure~\ref{fig:fig1}(a) illustrates the pump spectral shaping approach.
Starting from a broadband femtosecond laser pulse, a programmable pulse shaper (see details in App.~\ref{app:pulse shaper}) carves a multi-Gaussian spectrum from the original pulse envelope.
The corresponding multi-Gaussian PEF (see lower panel) exhibits the characteristic anti-diagonal structure in the signal-idler wavelength space, as imposed by energy conservation during the conversion.
Complementing this pump shaping, the use of symmetric group-velocity matching conditions in type-II KTP crystals with aperiodic poling patterns tailors a corresponding multi-Gaussian PMF~\cite{Morrison2022}.
Figure~\ref{fig:fig1}(b) reports the nonlinearity profile and corresponding poling pattern (depicted in the inset) of the engineered crystal (see details in App.~\ref{app:photon sources}) using a newly-developed sub-coherence-length algorithm~\cite{Frick}.
The specific poling design introduces a $\pi$-phase shift between adjacent antinodes of the PMF, as shown by simulations in the lower panel.
The resulting JSA, presented in Fig.~\ref{fig:fig1}(c), exhibits four well-separated and symmetrically arranged frequency bins.
The marginal distributions confirm no overlap between peaks of the multi-peak structure in both signal and idler wavelengths.
At the crystal's design pump central wavelength of \qty{791}{nm} (degenerate down-conversion at \qty{1582}{nm}), the Schmidt decomposition~\cite{Zielnicki2018} of the simulated JSA yields a Schmidt number $K=2.033$, indicating that the symmetric frequency-bin pattern is close to a two-mode maximally entangled state.

Figure~\ref{fig:fig1}(d) presents our experimental setup for hyperentanglement generation.
A mode-locked Ti:sapphire laser operating at a repetition rate of \qty{76}{MHz} and pulse duration of approximately \qty{150}{fs} serves as the pump source.
The pump pulses are spectrally shaped using a programmable pulse shaper based on a $4f$-configuration with a spatial light modulator placed in the Fourier plane~\cite{Faleo2025} (see details in App.~\ref{app:pulse shaper}) to produce the required multi-Gaussian spectrum.
The two Gaussian peaks are separated around the central wavelength to match the expected separation of the PMF peaks.
The shaped laser beam is split to pump two SPDC sources with a pump power of \qty{\approx 15}{mW} per source.
Each source consists of a custom aKTP crystal, with the design as presented in Fig.~\ref{fig:fig1}(b), placed within a Sagnac loop interferometer~\cite{Fedrizzi2007}.
This allows simultaneous generation of polarisation and frequency-bin entangled photon pairs~\cite{Morrison2022}, resulting in the hyperentangled state
\begin{equation}
\label{eq:hyperentangled state}
\begin{split}
    \ket{\psi} & = \ket{\psi_{\text{p}}} \otimes \ket{\psi_{\text{f}}}\\
    & = \frac{1}{2\sqrt{2}} \left( \ket{\text{HV}}_{\text{s,i}} + e^{i\varphi}\ket{\text{VH}}_{\text{s,i}}\right) \otimes \\
    &\left( \ket{\omega_1,\omega_2}_{\text{s,i}} -\ket{\omega_2,\omega_1}_{\text{s,i}} + \ket{\omega_2,\omega_3}_{\text{s,i}} -\ket{\omega_3,\omega_2}_{\text{s,i}} \right),
\end{split}
\end{equation}
with signal and idler modes ``s'' and ``i'', $\omega_1 > \omega_2 > \omega_3$ as the central frequencies of each frequency bin in the marginal distributions of Fig.~\ref{fig:fig1}(c), and $\varphi$ the relative phase set by the pump polarisation.
This hyperentangled state is the tensor product between a polarisation Bell state and a sum of two frequency Bell states $\ket{\psi^-_{\text{f}}}$, i.e., $\ket{\psi_{\text{f}}} \propto \ket{\psi^-_{\text{f},1}} + \ket{\psi^-_{\text{f},2}} \propto \left( \ket{\omega_1,\omega_2}_{\text{s,i}} -\ket{\omega_2,\omega_1}_{\text{s,i}} \right) + \left( \ket{\omega_2,\omega_3}_{\text{s,i}} -\ket{\omega_3,\omega_2}_{\text{s,i}} \right)$.
The polarisation state can be arbitrarily tuned to any other Bell state by in-fibre polarisation controllers. 
In order to match the dual-wavelength range of the Sagnac loop optics, the source is operated at a pump central wavelength of \qty{775.5}{nm} (degenerate down-conversion at \qty{1551}{nm}).
This results in a slight deviation from the ideal symmetry in the resulting JSA, yielding a simulated Schmidt number of $K_{\text{sim}}=2.088$.
Using the same design wavelength for the crystal and Sagnac loop optics would mitigate this issue.

The direct, full tomography of hyperentangled states comprising frequency-bin DOF requires electro-optic modulation~\cite{Lu2023_2} or nonlinear sum-frequency generation~\cite{Eckstein2011, Serino2023}.
Here, we demonstrate the hyperentanglement by combining spectral entanglement characterisation with polarisation-resolved TOFS and HOM interference.

\begin{figure*}
    \centering
    \includegraphics{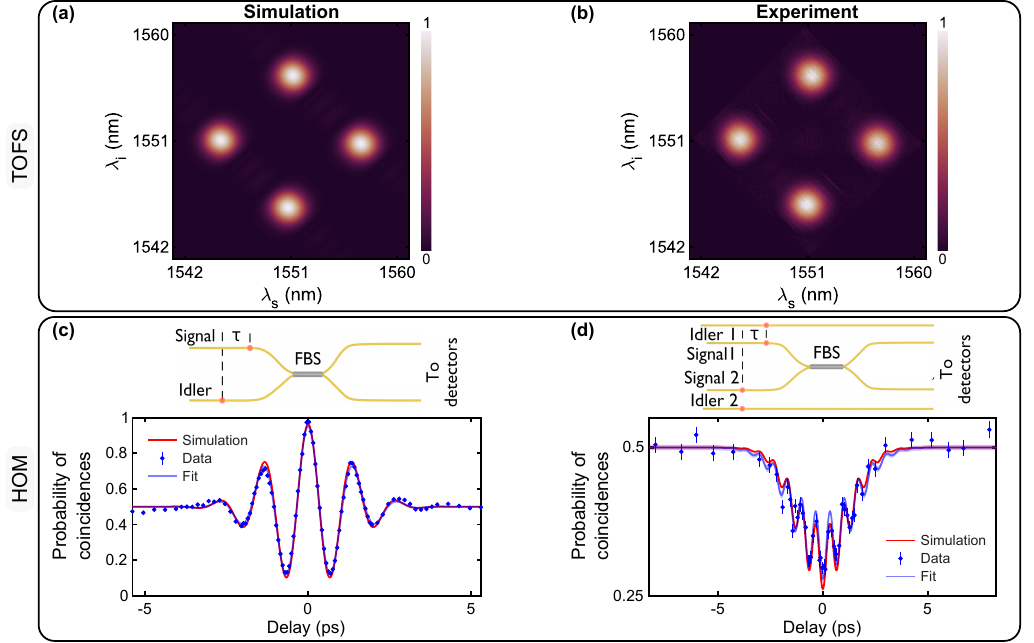}
    \caption{Spectral characterisation.
    \textbf{(a)}-\textbf{(b)} Simulated and experimental $\sqrt{\text{JSI}}$ obtained via TOFS, resulting in Schmidt numbers of $K_{\text{sim}}=2.088$ and $K_{\text{exp}}=2.098(2)$, respectively.
    \textbf{(c)} Intra-pair HOM interference at a fibre beam splitter (see upper inset).
    The simulation of the interference pattern assumes a JSA equivalent to $\sqrt{\text{JSI}}$ of panel (b) with a $\pi$-phase shift as shown in Fig.~\ref{fig:fig1}(c).
    The interference visibility of the simulation and of the fit are \qty{92.62(2)}{\percent} and \qty{90.3(4)}{\percent}, respectively. 
    \textbf{(d)} Heralded inter-pair HOM interference (see upper inset).
    The red line shows the simulation of the interference assuming JSAs as in (c).
    The blue fit line is accompanied by a shaded area representing its one-sigma uncertainty region.  
    The visibility of the simulation and fit are \qty{47.65(1)}{\percent} and \qty{44.2(1.2)}{\percent}, respectively.}
    \label{fig:fig3}
\end{figure*}

\section{\label{sec:spectral characterisation and hong-ou-mandel interference}Spectral characterisation}

We first validated the shaping of the joint spectrum and confirmed its spectral entanglement in the absence of polarisation entanglement.
This is done using two methods: TOFS~\cite{Avenhaus2009} to measure the joint spectral intensity (JSI) and its Schmidt number, and HOM interference to witness spectral entanglement via anti-bunching peaks~\cite{Fedrizzi2009} and to confirm the expected phase structure in the JSA as presented in Fig.~\ref{fig:fig1}(c).

Figure~\ref{fig:fig3}(a) shows the simulated $\sqrt{\text{JSI}}$, calculated from the designed PEF and PMF, yielding a Schmidt number of $K_{\text{sim}} = 2.088$.
The experimental $\sqrt{\text{JSI}}$, measured via TOFS (see App.~\ref{app:detection and tofs} for details) and shown in Fig.~\ref{fig:fig3}(b), agrees with the simulation with a Schmidt number of $K_{\text{exp}} = 2.098(2)$.
The uncertainty of this and all following experimental results as well as JSI-based simulations is computed with Monte Carlo simulations using Poissonian counting statistics in the measured JSI.

We performed HOM interference measurements using unidirectional pumping within the Sagnac loop to erase the entanglement in polarisation.
Figure~\ref{fig:fig3}(c) shows the intra-pair HOM interference, where signal and idler photons from the same pair are interfered at a fiber beam splitter (FBS) as their relative delay $\tau$ is scanned.
The experimental data (uncertainty bars representing the Poissonian counting statistics) exhibit a spatial anti-bunching pattern, here with a global maximum at zero delay, resulting in a fitted interference visibility of \qty{90.3(4)}{\percent} (see App.~\ref{appendix:intra pair fit model} for details of the fit model).
Given the absence of polarisation entanglement, this anti-bunching behaviour is indicative of frequency-bin entanglement~\cite{Fedrizzi2009}, and the obtained visibility quantifies the degree of spectral entanglement of the down-converted photon pair.
The measured visibility is close to the simulated value of \qty{92.62(2)}{\percent}, obtained from the theoretical HOM interference pattern in the absence of experimental imperfections such as unbalanced splitting ratios, multi-pair emission, and residual polarisation distinguishability (see details in App.~\ref{appendix:intra pair fit model}).
This and the following simulations are based on the JSA reconstructed from the measured $\sqrt{\text{JSI}}$ assuming the design $\pi$-phase shift between adjacent frequency bins, as shown in Fig.~\ref{fig:fig1}. Note that a zero phase shift in the PMF would produce a bunching dip at zero delay (see App.~\ref{appendix:intra pair fit model}).
The obtained visibility values are primarily limited by the use of the crystal with a pump wavelength that considerably differs from the designed one, as mentioned in Sec.~\ref{sec:joint spectral amplitude shaping}.
In fact, simulations at the design wavelength, as reported in Fig.~\ref{fig:fig1}, result in a theoretical visibility of up to \qty{\approx 99.9}{\percent}.

To verify that no additional spectral phase correlation besides the designed phase structure is present in the JSA, we performed heralded inter-pair HOM interference between signal photons from two independent sources, as shown in Fig.~\ref{fig:fig3}(d). 
The data closely follow the rapid oscillations modulated by a Gaussian envelope of the interference pattern predicted by the simulation based on the measured JSI in Fig.~\ref{fig:fig3}(b).
The expected simulated interference visibility is \qty{47.65(1)}{\percent}, and the fitted interference visibility is \qty{44.2(1.2)}{\percent} (see App.~\ref{appendix:inter pair fit model} for details).
Considering the aforementioned experimental imperfections and potential fabrication differences between the crystals of the two sources, we conclude that no major intra-bin spectral phase correlations are present in the JSAs and the overall phase structure is as designed.
A minor degradation of the HOM-visibility can arise from group delay dispersion in the pump spectral shaping setup, see Ref.~\cite{Faleo2025} for a more detailed analysis.

\section{\label{sec:spectral shaping tunability}Spectral Shaping Tunability}

\begin{figure}
    \centering
    \includegraphics{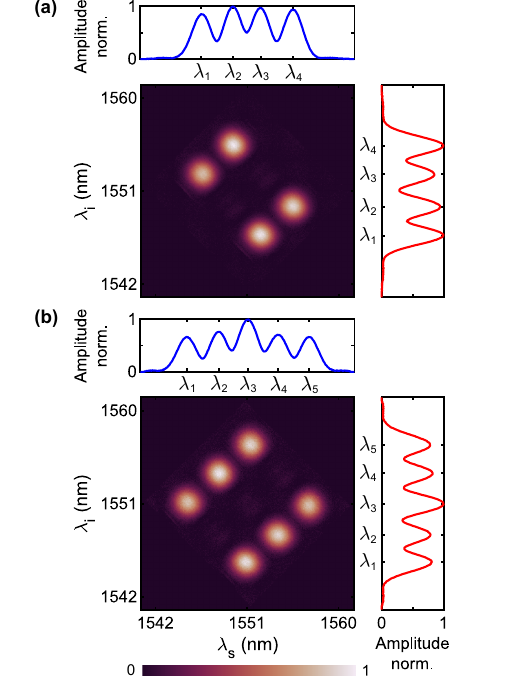}
    \caption{Spectral shaping tunability. \textbf{(a)}-\textbf{(b)} Experimental $\sqrt{\text{JSI}}$ and marginal distributions for a double and triple Gaussian shaping, respectively, with a peak spacing equal to half that of the Gaussian PMF.
    (a) Schmidt decomposition results in a Schmidt number of $K=3.881(2)$.
    (b) Schmidt decomposition results in a Schmidt number of $K=3.701(2)$. }
    \label{fig:fig4}
\end{figure}

A key feature of our approach is the ability to reconfigure the frequency-bin structure by modifying the pump spectral shaping while using the same engineered crystals.
Figure~\ref{fig:fig4} demonstrates this tunability by showing experimental $\sqrt{\text{JSI}}$ measurements for two different pump configurations, both ideally leading to a Schmidt number $K=4$.

\begin{figure*}
    \centering
    \includegraphics{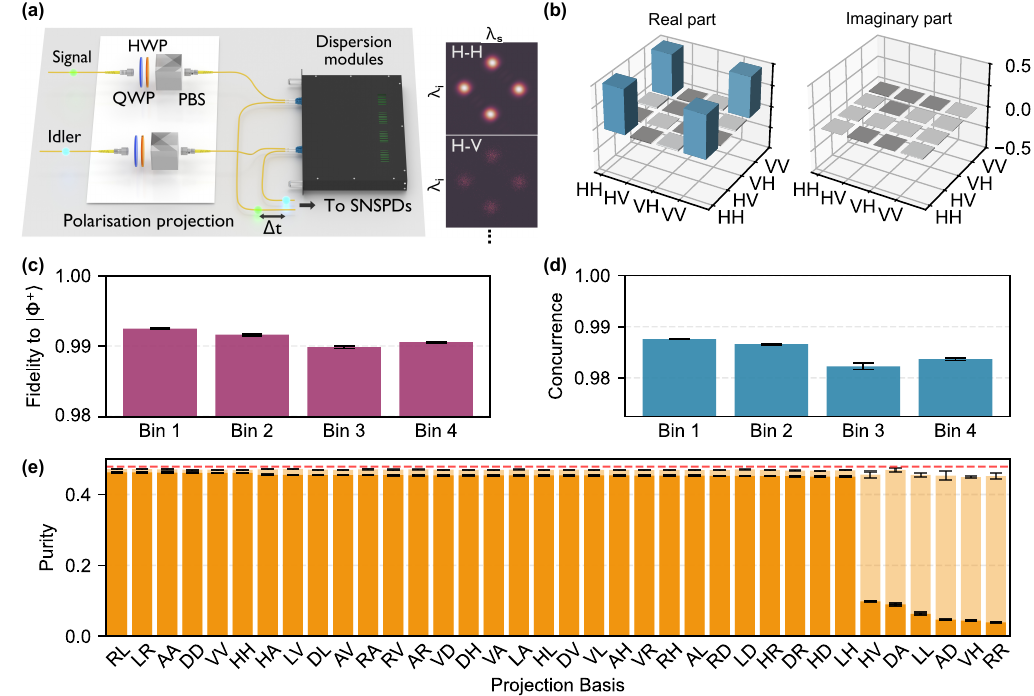}
    \caption{Polarisation entanglement of individual frequency bins. \textbf{(a)} Photons pairs undergo a polarisation projection by employing a quarter-wave plate (QWP), HWP, and PBS combination.
    At the output, the photons propagate through dispersion modules and, for each projection, coincidence events are measured with SNSPDs to perform a TOFS measurement and reconstruct the JSI (see examples for H-H and H-V projections).
    Polarisation state tomography for each frequency bin is performed by using a set of 36 such polarisation projections.
    \textbf{(b)} Real and imaginary parts of the reconstructed density matrix in the polarisation DOF via polarisation-resolved TOFS for one of the four frequency bins (``Bin 1'' in the subsequent panels).
    \textbf{(c)} Fidelities of the reconstructed polarisation states with respect to the target Bell state $\ket{\phi^+_{\text{p}}}$ of the four individual frequency bins.
    \textbf{(d)} Concurrence of the reconstructed polarisation-entangled states of the four individual frequency bins.
    \textbf{(e)} Spectral purity of the measured $\sqrt{\text{JSI}}$ for each of the required 36 polarisation projections. The red dashed line is the value obtained via simulation, $P_{\text{sim}}=K_{\text{sim}}^{-1}=0.4789$, dark orange bars are the raw measured values, and the light orange bars are the measured values corrected for counting statistic noise.}
    \label{fig:fig5}
\end{figure*}

Figure~\ref{fig:fig4}(a) shows the result for a double Gaussian pump shaping with peak spacings equal to half that of the Gaussian PMF.
This removes the $C_4$-rotational symmetry of the JSI in Fig.~\ref{fig:fig3}(b), producing a frequency-bin pattern with four spectral modes associated with two frequency-bin Bell states, i.e., $\ket{\psi_{\text{f}}} \propto \left( \ket{\omega_1,\omega_3}_{\text{s,i}} -\ket{\omega_3,\omega_1}_{\text{s,i}} \right) + \left( \ket{\omega_2,\omega_4}_{\text{s,i}} -\ket{\omega_4,\omega_2}_{\text{s,i}} \right)$, where $\omega_{j}$ are the central frequencies of the four modes $j=1,\dots,4$ (see marginals insets).
Schmidt decomposition of the simulated JSA and measured $\sqrt{\text{JSI}}$ yields $K_{\text{sim}} = 3.882$ and $K_{\text{exp}} = 3.881(2)$, respectively, indicating a near maximally entangled four-mode state.
The main limitation to reaching the ideal value of $K=4$ arises from the partial overlap among the frequency bins, as evidenced by the marginal distributions, which can be removed, for example, by starting from a larger spacing design between the Gaussian peaks of the PMF.

Figure~\ref{fig:fig4}(b) presents the result for a triple Gaussian shaping, whose resulting frequency state is associated with the sum of a Bell state and a qutrit state $\ket{\psi_{\text{f}}} \propto \left( \ket{\omega_2,\omega_4}_{\text{s,i}} -\ket{\omega_4,\omega_2}_{\text{s,i}} \right) + \left( \ket{\omega_3, \omega_1}_{\text{s,i}} - \ket{\omega_3, \omega_5}_{\text{s,i}} + \ket{\omega_1, \omega_3}_{\text{s,i}} - \ket{\omega_5, \omega_3}_{\text{s,i}}  \right)$.
The measured Schmidt number is $K = 3.701(2)$, and simulations lead to a Schmidt number of $K = 3.600$.
The difference between simulation and experiment is mainly due to slightly uneven peak heights in the joint spectrum.
The marginal distributions confirm the modified structure, in both cases, and the presence of partial overlap among the frequency bins (here highlighted by considering the $\sqrt{\text{JSI}}$ instead of the JSI), which reduces the Schmidt number from the ideal value of $K = 4$.

\section{\label{sec:polarisation entanglement in individual frequency bins}Polarisation Entanglement in Individual Frequency Bins}

\begin{figure*}
    \centering
    \includegraphics{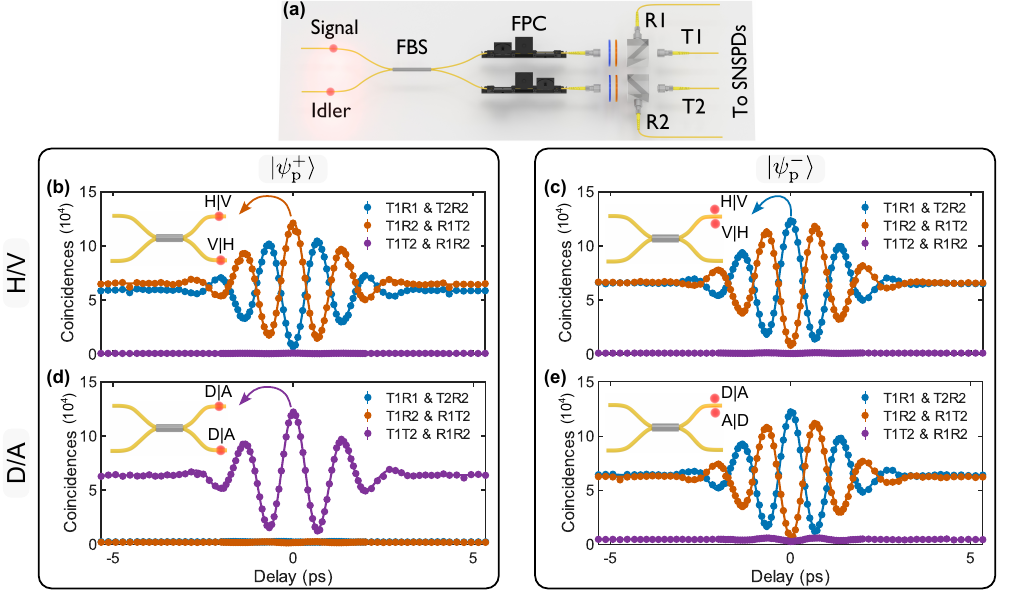}
    \caption{Exchange symmetry of the hyperentangled photon pairs.
    \textbf{(a)} Photon pairs are interfered by scanning their temporal delay at the FBS. 
    At each output, tomography stages perform a polarisation projection and two-fold coincidence counts between the four output ports (T1, R1, T2, R2) are recorded over \qty{20}{s} with SNSPDs.
    \textbf{(b)}-\textbf{(e)} Each panel displays three coincidence combinations: cross-polarisation-outputs (T1R1 \& T2R2 in blue, T1R2 \& R1T2 in orange) and same-polarisation-outputs (T1T2 \& R1R2 in purple).
    The solid lines are the fits of the data (dots).
    The insets illustrate the beam splitter outputs at zero delay (highlighting bunching or antibunching behaviour) and the possible photons' polarisations.
    (b)-(c) In the H/V basis, both Bell states produce T-R coincidences since $\ket{\text{HV}}\pm\ket{\text{VH}}$ always pairs orthogonal polarisations.
    (d)-(e) In the D/A basis, $\ket{\psi^+_{\text{p}}}$ transforms to $\propto\ket{\text{DD}}+\ket{\text{AA}}$, yielding same-polarisation coincidences (purple trace in (d)), while $\ket{\psi^-_{\text{p}}}$ transforms to $\propto\ket{\text{AD}}-\ket{\text{DA}}$, yielding again cross-polarisation coincidences as in (c).}
    \label{fig:fig6}
\end{figure*}

Having established the frequency-bin structure, we characterised the polarisation entanglement of each bin generated by bidirectionally pumping the crystal in the Sagnac interferometer.
To do so, we employed polarisation-resolved TOFS as shown in Fig.~\ref{fig:fig5}(a).
The signal and idler photons from each source first pass through polarisation-tomography stages and are then sent through dispersion modules to perform TOFS.
We reconstruct the JSI in a polarisation-resolved manner by recording coincidences over one hour for 36 polarisation projections (total integration time of 36 hours) with superconducting nanowire single-photon detectors (SNSPDs) and a time tagging unit.
The set of 36 polarisation projection measurements provides access to polarisation state tomography for each frequency bin of the JSI (see App.~\ref{appendix:tomography}), while also enabling the verification of the JSI's overall Schmidt number.

Panels (b)-(e) of Fig.~\ref{fig:fig5} report the results of this analysis, where the target maximally polarisation-entangled Bell state, set by the FPCs, is $\ket{\phi^+_{\text{p}}}=(\ket{\text{HH}}+\ket{\text{VV}})/\sqrt{2}$.
Figures~\ref{fig:fig5}(b) shows the real and imaginary parts of the reconstructed density matrix for one frequency bin (``Bin 1" in the later panels; see Fig.~\ref{fig:appendix_tomography} in App.~\ref{appendix:tomography} for the bin labelling), which are consistent with the density matrix expected for the Bell state $\ket{\phi^+_{\text{p}}}$.
This is confirmed by the calculated fidelities of the reconstructed polarisation states with respect to the target state $\ket{\phi^+_{\text{p}}}$, for all four frequency bins, as presented in Figure~\ref{fig:fig5}(c).
Fidelities equal or exceed \qty{99}{\percent}, with values of \qty{99.248(4)}{\percent}, \qty{99.16(4)}{\percent}, \qty{98.98(4)}{\percent}, and \qty{99.053(9)}{\percent} for Bins 1 through 4, respectively.
Figure~\ref{fig:fig5}(d) shows the concurrence, with values of \qty{98.76(3)}{\percent}, \qty{98.66(3)}{\percent}, \qty{98.23(3)}{\percent}, and \qty{98.37(3)}{\percent}, attesting highly polarisation-entangled states.

To verify that the frequency-bin entanglement is maintained independently of the polarisation measurement basis, we performed the Schmidt decomposition of the measured $\sqrt{\text{JSI}}$ for all 36 polarisation projections.
Figure~\ref{fig:fig5}(e) displays the purity $P = K^{-1}$ for each projection and the limit of the simulated purity $P_{\text{sim}} = K_{\text{sim}}^{-1} = 0.4789$.
The Schmidt decomposition is highly dependent on the counting statistics.
For projections where only very few counts are recorded, the number of Schmidt modes required to decompose the joint spectrum increases, resulting in a reduction of the measured purity~\cite{PsiQuantum2025, Faleo2025}.
This effect is clearly visible in the raw measured values (dark orange bars) corresponding to projection bases where zero counts are expected for an input $\ket{\phi^+_{\text{p}}}$ state (see for example HV projection in the inset of Fig.~\ref{fig:fig5}(a)).
To overcome this limitation, we applied a noise correction procedure similar to that reported in Ref.~\cite{PsiQuantum2025} \footnote{In this case, we obtained the noiseless joint spectrum using the first four singular values of the Schmidt decomposition, rather than employing only the highest one.}, which shows a significant restoration of the purity values (light orange bars).
The average corrected purity across all projections is $0.468(1)$, corresponding to a Schmidt number of $2.139(5)$.
Excluding the six projections with low counting statistics results in a purity of $0.4699(2)$ and Schmidt number of $2.1282(9)$.
This consistent purity across all projections, which agrees with the results in Fig.~\ref{fig:fig3}, indicates that the frequency-bin structure is preserved regardless of the polarisation measurement basis, as expected for a hyperentangled state.

\section{\label{sec:HOM hyperentangled state}Symmetry of the Hyperentangled Photon Pairs}

To demonstrate that entanglement is simultaneously present in both DOF and not a mere mixture of states, we analyse how the exchange symmetry properties of the hyperentangled state manifest in polarisation-resolved HOM interference, as depicted in Fig.~\ref{fig:fig6}(a).
Hyperentangled photon pairs are interfered at a polarisation-insensitive fibre beam splitter and tomography modules perform a polarisation projection at the two outputs.
Both the transmitted (T) and reflected (R) outputs of each tomography stage (labelled as ``T1/R1" at stage ``1" and respectively at stage ``2") are connected to SNSPDs in order to record all combinations of two-fold coincidence counts between the FBS output ports.
By switching between projections in the H/V and D/A bases of the tomography stages and from the complete set of two-fold combinations among the four outputs, we can simultaneously distinguish between bunching or antibunching at the beam splitter, as well as between the presence of a symmetric ($\ket{\psi^+_{\text{p}}}$, $\ket{\phi^+_{\text{p}}}$, $\ket{\phi^-_{\text{p}}}$) or antisymmetric ($\ket{\psi^-_{\text{p}}}$) polarisation state.

The state of the hyperentangled photon pairs is a product of the polarisation and frequency components, and the exchange symmetry of the combined state determines the interference behaviour.
The central antibunching peak in Fig.~\ref{fig:fig3}(c) arises because the frequency-bin entangled state is antisymmetric under particle exchange, while the total two-photon state, including the spatial modes, needs to be symmetric (as required for bosons).  
Moreover, the polarisation Bell states have definite exchange symmetry: $\ket{\psi^+_{\text{p}}} = \left( \ket{\text{HV}}\ + \ket{\text{VH}} \right)/\sqrt{2}$ is symmetric, while $\ket{\psi^-_{\text{p}}} = \left( \ket{\text{HV}}\ - \ket{\text{VH}} \right)/\sqrt{2}$ is antisymmetric.
Therefore, the use of a $\ket{\psi^+_{\text{p}}}$ or a $\ket{\psi^-_{\text{p}}}$ state necessarily leads to an antibunched or a bunched interference pattern, respectively (see also App.~\ref{appendix:polarisation resolved HOM}).

Panels (b) to (e) of Fig.~\ref{fig:fig6} report the results of these measurements when setting either a $\ket{\psi^+_{\text{p}}}$ (left column) or $\ket{\psi^-_{\text{p}}}$ polarisation-state (right column) by rotating the HWP at the input of the Sagnac loop by \qty{45}{\degree}.
In addition, the polarisation is measured either in the H/V (top row) or D/A basis (bottom row).
The measured coincidence traces in all four panels exhibit the characteristic beating pattern arising from the frequency-bin entanglement (cf. Fig.~\ref{fig:fig3}(c)), whereas a mixture of frequency-bin occupations would produce a flat trace.
Comparing the measurements in the H/V basis of panels (b) and (c), we observe that the interference pattern inverts when switching between the two Bell states.
Specifically, the coincidences between cross-polarisation outputs at different FBS ports (T1R2 \& R1T2, shown in orange) show an antibunching peak for $\ket{\psi^+_{\text{p}}}$ but a bunching dip for $\ket{\psi^-_{\text{p}}}$, and the opposite relation for cross-polarisation outputs at the same FBS port (T1R1 \& T2R2, shown in blue).
This inversion and the absence of coincidences in same-polarisation outputs (T1T2 \& R1R2, shown in purple; they vanish for $\psi_{\text{p}}$ states) directly reflect the different symmetry properties of the two polarisation-entangled states.
Crucially, the antibunching observed with the state $\ket{\psi^+_{\text{p}}}$ requires the frequency state to be entangled too, and to have an antisymmetric exchange symmetry, as does the bunching pattern obtained with the state $\ket{\psi^-_{\text{p}}}$.
This interpretation can be corroborated by projection measurements in the D/A basis, as presented in panels (d) and (e).
While in the H/V basis both states have the form $\ket{\text{HV}}\pm\ket{\text{VH}}$, meaning each photon pair always exits in the cross-outputs T and R, in the D/A basis their form differs.
The symmetric state becomes $\ket{\psi^+_{\text{p}}}=(\ket{\text{DD}}+\ket{\text{AA}})/\sqrt{2}$, where both photons share the same polarisation and always exit through the same port, either transmitted or reflected, as visible in panel (d).
In contrast, the antisymmetric state transforms to $\ket{\psi^-_{\text{p}}}=(\ket{\text{AD}}-\ket{\text{DA}})/\sqrt{2}$, resulting in a behaviour that is identical to that observed in the H/V basis, as confirmed by panel (e).

The observed average visibilities for the measurements in the H/V basis, excluding same-polarisation coincidences, are \qty{88.1(8)}{\percent} and \qty{88.0(8)}{\percent} for $\ket{\psi^-_{\text{p}}}$ and $\ket{\psi^+_{\text{p}}}$, respectively.
For the D/A basis, we obtained a visibility of \qty{91.3(6)}{\percent} for $\ket{\psi^+_{\text{p}}}$ (T1T2 \& R1R2 trace) and an average visibility of \qty{91.0(9)}{\percent} for $\ket{\psi^-_{\text{p}}}$ (only cross-polarisation coincidences).
These results are in line width the measurement in Fig.~\ref{fig:fig3}(c), with deviations that are mainly ascribable to differences in the optimisation of the FPCs.

The simultaneous observation of frequency beating and bunching/antibunching behaviour dependent on the symmetry of the polarisation state provides unambiguous evidence that the photon source generates hyperentangled states with
correlations in both polarisation and frequency-bin DOF.
In doing so, we have also demonstrated the possibility of acting on one DOF independently, leaving the other one undisturbed.

\section*{Discussion}

Our results demonstrate the generation of photon pairs hyperentangled in polarisation and frequency-bin DOF.
The combination of programmable pump spectral shaping and crystal engineering provides independent control over the PEF and PMF, enabling precise tailoring of the JSA and, therefore, of the two-dimensional frequency-bin structure.
The experimentally achieved Schmidt number of \qty{2.098(2)}{\percent} in the rotation-symmetric four-bin scenario of Fig.~\ref{fig:fig3}(b) indicates a two-mode frequency-bin entangled state.
The Schmidt numbers for two additional pump configurations, as shown in Fig.~\ref{fig:fig4}, demonstrate two distinct methods of accessing four-mode frequency-bin entangled states.
Such tunability is a key advantage of the here presented approach: the frequency-bin structure and dimensionality can be reconfigured by reprogramming the pulse shaper without modifying the crystal.
This flexibility enables adaptation to different protocol requirements.

The HOM interference measurements of polarisation-separable photon pairs provide additional validation to the JSI reconstruction.
The antibunched intra-pair interference pattern of Fig.~\ref{fig:fig3}(c) and its visibility reflect the quality of the frequency-bin entanglement and the phase relationships between the frequency bins.
The inter-pair interference [Fig.~\ref{fig:fig3}(d)] demonstrates the absence of major unwanted spectral phase correlations in the JSA and the potential for multi-photon protocols involving independent sources.
Finally, the frequency-resolved state tomography of Fig.~\ref{fig:fig5} and the HOM interference of hyperentangled photon pairs, as shown in Fig.~\ref{fig:fig6}, confirm the entanglement in both DOF and their independence, as it is possible to tune the polarisation component to any Bell state without affecting the frequency component.

The high fidelities ($>$\qty{99}{\percent}) and concurrences ($>$\qty{98}{\percent}) achieved for polarisation entanglement across all frequency bins confirm that both DOF can be simultaneously exploited as quantum resources.
This is particularly relevant for quantum information processing protocols that can leverage hyperentanglement to enhance their capacity, noise resilience or efficiency.
The telecom wavelength operation ensures compatibility with existing fibre-optic infrastructure.

The current work can be extended in several directions.
For example, the frequency-bin dimensionality could be further increased, giving access to promising quantum states~\cite{Fabre2020, Drago2022}.
A source of an in-line eight-bin joint spectrum has been obtained by solely shaping the nonlinearity~\cite{Morrison2022}.
With a sufficiently broad pump laser spectrum (pulse duration in the order of tens of fs), the approach presented here could be used to extend this in-line configuration to a two-dimensional, symmetric $8 \times 8$ grid of 32 independent frequency bins (Schmidt modes).   
This frequency-bin structure is naturally suited for wavelength-division multiplexed quantum networks, where different frequency bins could be routed to different nodes in multi-user anonymous protocols~\cite{Huang2022_qkd}.
The hyperentanglement photon pairs could be employed for single-copy entanglement distillation~\cite{Ecker2021, Chen2025} and faithful distribution of entanglement over noisy channels~\cite{Ecker2022}.
Furthermore, generation of entanglement in additional DOF, such as time-bin~\cite{Xavier2025} or pulse-mode~\cite{Chiriano2023}, can be used to create even higher-dimensional hyperentangled quantum states and resource states for cluster states generation~\cite{Reimer2019}.

\section*{Acknowledgements}

\noindent The authors acknowledge Alessandro Fedrizzi and Christopher L. Morrison for helpful discussions.
This research was funded in part by the Austrian Science Fund (FWF) projects 10.55776/FG5, 10.55776/F71, 10.55776/W1259, 10.55776/COE1, 10.55776/TAI556, 10.55776/PIN6357923, 10.55776/PIN9593224,  and infrastructure funding from FFG (grant no. FO999896024). This research was partially funded by Förderkreis 1669 of the University of Innsbruck.

\bibliography{references.bib}

\appendix

\section{\label{appendix:experimental_setup}Details experimental setup}
\begin{figure*}
    \centering
    \includegraphics[width=\linewidth]{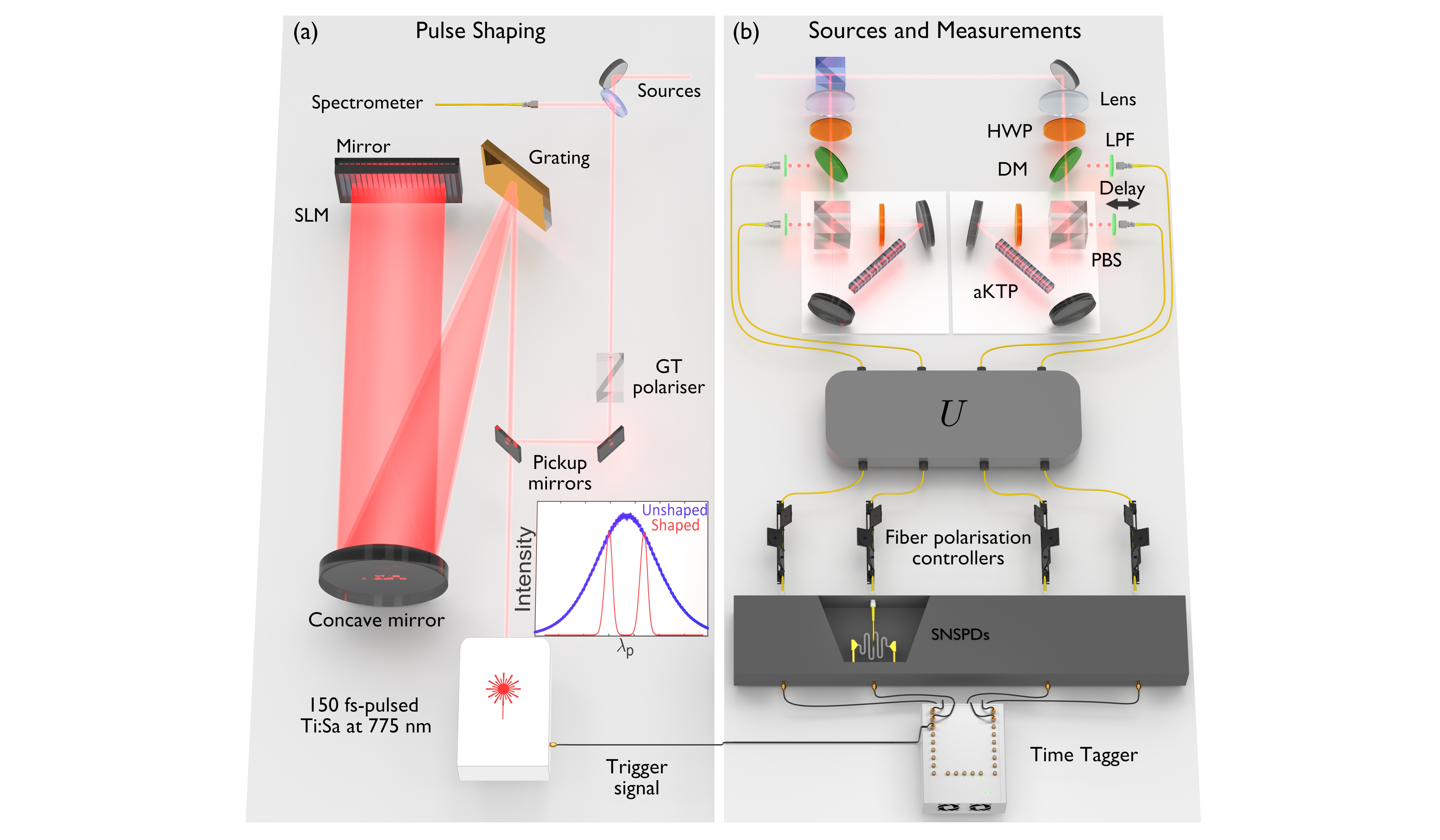}
    \caption{Detailed experimental setup.
    \textbf{(a)} The pump laser spectrum is shaped using a folded 4$f$ pulse shaper. A reflective diffraction grating and concave mirror map frequency components onto a spatial light modulator (SLM) at the Fourier plane, which imposes a wavelength-dependent polarisation rotation. The beam is retro-reflected by a planar mirror with a slight vertical tilt, recomposed at the grating, and separated from the input beam via pickup mirrors. A Glan-Taylor (GT) polariser converts the polarisation modulation into amplitude shaping. Part of the output is monitored on a spectrometer for closed-loop optimisation of the target spectrum. 
    \textbf{(b)} The shaped beam is split and focused into two SPDC sources, each consisting of a custom aKTP crystal inside a Sagnac loop as described in Fig.~\ref{fig:fig1}(d). Depending on the measurement, photons undergo a transformation $U$ corresponding to either polarisation-resolved or non-polarisation-resolved TOFS or HOM interference configurations (including the FPCs of Fig.~\ref{fig:fig1}(d)). Detection is performed with superconducting nanowire single-photon detectors (SNSPDs), and events are recorded using a time-tagging system.}
    \label{fig:appendix_setup}
\end{figure*}

The detailed representation of the setup is reported in Fig.~\ref{fig:appendix_setup} and discussed in the following sections.

\subsection{\label{app:pulse shaper}Pulse Shaping}
Fig.~\ref{fig:appendix_setup}(a) shows the pump spectral shaping.
The laser beam enters a $4f$ pulse shaper in a folded configuration~\cite{Monmayrant2010}.
This is composed of a diffraction grating (groove density \qty{2400}{lines/mm}), a concave mirror with focal length of \qty{500}{mm}, and a spatial light modulator with 128 liquid crystal pixels equipped with a planar mirror on the back (reflecting the beam slightly upwards).
Pickup mirrors intercept the outgoing laser beam vertically displaced at the output of the $4f$-setup and redirect it to the sources.

As described in Ref.~\cite{Faleo2025}, this setting allows a fine-tuning of the laser spectrum.
The grating-concave mirror pair maps the laser frequency components into different positions along the transverse direction in the Fourier plane, where the SLM and planar mirror sit.
The pixels are made of birefringent liquid crystals that can be rotated by properly tuning the applied electric voltage. 
Each pixel therefore acts as a waveplate imposing a polarisation rotation to a specific set of frequencies $\delta\nu$ centred around a frequency $\bar{\nu}$, where $\bar{\nu}$ is a function of the position along the horizontal direction of the transverse plane. 
At the output, a Glan-Taylor polariser shapes the spectrum by reducing the intensity of the different frequency components $\bar{\nu} \pm \delta\nu / 2$ depending on the polarisation rotation imposed by the SLM.
To achieve a target output spectrum, optimisation of the voltage levels is performed with a feedback loop by measuring the shaped spectrum with a spectrometer. 

\subsection{\label{app:photon sources}Photon Sources}

Each polarisation-entangled photon pair source consists of an aKTP crystal of length $\text{L}=\qty{30}{mm}$ placed at the centre of a Sagnac loop interferometer, as shown in Fig.~\ref{fig:appendix_setup}(b).
The pump laser is focused to reach a beam waist of \qty{\approx 66}{\mu\meter} at the position of the crystals, while the waist associated with the output collection optics is \qty{\approx 68.5}{\mu\meter} (see also Ref.~\cite{Faleo2025}).

The crystal is designed using sub-coherence-length domain engineering by tracking the ideal nonlinearity profile, which best approximates the target multi-Gaussian PMF~\cite{Graffitti2017, Frick}.
In our case, the PMF required for a double-Gaussian profile corresponds to
\begin{multline}
\label{eq:PMF Delta k}
    \phi(\Delta k) = \frac{1}{\sqrt{2 \pi } \xi } \biggl[ \text{exp}\left(-\frac{\left(\Delta k - \Delta k_0 -\frac{\epsilon }{2}\right)^2}{2 \xi ^2}\right)\\
    - \text{exp}\left(-\frac{\left(\Delta k - \Delta k_0 + \frac{\epsilon }{2}\right)^2}{2 \xi ^2}\right) \biggr],
\end{multline}
where $\Delta k$ is the (wavelength-dependent) phase mismatch, $\Delta k_0$ the phase mismatch at the degenerate design wavelengths, and the separation and width of the Gaussian peaks in the wavevector space are determined by $\Delta k_0 \pm \epsilon/2$ and $\xi$, respectively.
The PMF and the associated nonlinearity profile $g(z)$ along the crystal axis $z$ are linked by a Fourier transform.
Therefore, the desired nonlinearity profile that results in Eq.~\eqref{eq:PMF Delta k} equals
\begin{align}
    g(z) = i \sqrt{\frac{2}{\pi}} \sin{\left(\frac{\epsilon z}{2}\right)}\text{exp}\left( i \text{$\Delta$k}_0 z -\xi^2 z^2/2 \right).
\end{align}
Figure~\ref{fig:fig1}(b) shows the absolute value of the nonlinearity profile for the crystals used in this work, designed to work with a pump wavelength of \qty{791}{nm} (main poling period \qty{\approx 23}{\micro\meter}, minimum domains size of \qty{9}{\micro\meter}, and maximum domains size of \qty{792.75}{\micro\meter}), with Gaussian widths of $\xi=4/\text{L}$, and separation $\epsilon\qty{\approx 1331}{m^{-1}}$.
The chosen Gaussian width $\xi$ is a compromise between the intra-bin spectral purity, with heralded photons from any one bin in Figure~\ref{fig:fig1}(c) having a theoretical spectral purity of \qty{\approx 99.3}{\percent} (see also Ref.~\cite{Faleo2025} for a comparison between different Gaussian width values), source brightness, and bandwidth of the resulting frequency bins.

The Sagnac loops are composed of a dual-wavelength (\qty{\approx 775}{nm} and \qty{\approx 1550}{nm}) polarisation beam splitter, a half-wave plate, and mirrors.
The half-wave plate at the input enables the pump laser to propagate in either a clockwise or counter-clockwise direction, or in both directions inside the Sagnac loop.
A balanced, bidirectional pumping configuration ideally produces maximally entangled pairs in the polarisation DOF~\cite{Fedrizzi2007,Meraner2021,Faleo2024}.

\subsection{\label{app:detection and tofs}Detection and TOFS}
The down-converted photons are collected in single-mode fibres (\qty{\approx 40}{\percent} heralding efficiency, including detector efficiency and fiber mating losses), undergo a transformation (``\emph{U}" in Fig.~\ref{fig:appendix_setup}(b)), and single-photon events are measured at each output with SNSPDs.
These have an average (bare) detection efficiency of \qty{\approx 85}{\percent} at \qty{1550}{nm}.
A time-tagger unit records single-photon events, as well as coincidence events among the four channels (two-fold, three-fold, and four-fold coincidences).
For TOFS and polarisation-resolved TOFS measurements (see Fig.~\ref{fig:fig5}(a)), the time-tagger unit collects three-fold coincidences in two-dimensional time histograms between a trigger channel and the signal and idler photon channels.
The trigger channel is obtained by using a fast photodiode detecting the train of pulses of the pump laser output.

In TOFS measurements, the dispersion modules provide a negative dispersion of \qty{-1350}{ps/nm}.
They are originally designed for commercial use to compensate for dispersion of \qty{\approx 80}{km} in fibres at telecommunications wavelengths.
This maps photon wavelengths to arrival times, enabling spectral resolution through timing measurements~\cite{Avenhaus2009}.
The combined time jitter is obtained by adding in quadrature the independent jitter contributions and is estimated to be \qty{\approx 37}{ps}.
This includes contributions from the SNSPDs, the time tagger unit, and the fast photodiode.
The combined time jitter negligibly affects the measurement as the time bin width used for the histograms is set between \qtyrange{80}{100}{ps}, considerably larger than the jitter value~\cite{Zielnicki2018}. 
JSI reconstruction is performed by applying the inverse wavelength-to-time mapping.

\section{\label{appendix:fit model}HOM interference}

With our spectral shaping setup, the pump envelope function can be well approximated with a double Gaussian function
\begin{multline}
\label{eq:PEF}
    \alpha(\omega_{\text{s}} + \omega_{\text{i}}) = \exp \left(-\frac{(\omega_{\text{i}} + \omega_{\text{s}} -(\omega_{\text{p}}-\delta ))^2}{2 \sigma^2} \right)  \\
    + \exp\left( -\frac{(\omega_{\text{i}} + \omega_{\text{s}}-(\omega_{\text{p}}+\delta) )^2}{2 \sigma^2} \right)
\end{multline}
where $\omega_{\text{s}}$ and $\omega_{\text{i}}$ are the signal and idler photon frequencies, and $\omega_{\text{p}} \pm \delta$ are the central frequencies of the Gaussians carved from the unshaped pump spectrum at a distance $\delta$ from the pump central frequency $\omega_{\text{p}}$.
The engineered phase-matching function can also be well approximated as a sum of two Gaussians
\begin{multline}
\label{eq:PMF}
    \phi(\omega_{\text{s}}, \omega_{\text{i}}) = \exp \left( -\frac{(\omega_{\text{s}} - \omega_{\text{i}} - \delta )^2}{2 \sigma^2} \right) \\
    - \exp \left(-\frac{(\omega_{\text{s}} - \omega_{\text{i}} + \delta )^2}{2 \sigma^2} \right)
\end{multline}
with the same width and separation as the pump envelope function, and with opposite signs as a consequence of the used poling pattern.
The corresponding joint spectral amplitude, up to a normalisation factor, is defined as
\begin{equation}
    f(\omega_{\text{s}}, \omega_{\text{i}}) \propto \alpha(\omega_{\text{s}} + \omega_{\text{i}}) \phi(\omega_{\text{s}}, \omega_{\text{i}}),
\end{equation}
and the state of the down-converted photon-pair (neglecting multi-pair production) is
\begin{equation}
    \ket{\psi_{\text{spdc}}}_{\text{s,i}}=\int \int f(\omega_{\text{s}}, \omega_{\text{i}}) a^{\dagger}(\omega_{\text{s}}) b^{\dagger}(\omega_{\text{i}}) \ket{0}_{\text{s,i}} \mathrm{d}\omega_{\text{s}} \mathrm{d}\omega_{\text{i}},
\end{equation}
where $a^{\dagger}(\omega_{\text{s}})$ and $b^{\dagger}(\omega_{\text{i}})$ are the bosonic creation operators generating a signal and idler photon in spatial modes $a$ and $b$, and at frequencies $\omega_{\text{s}}$ and $\omega_{\text{i}}$, respectively.

\subsection{\label{appendix:intra pair fit model}Intra-pair interference}

The coincidence probability as a function of the relative time delay $\tau$ between photons arriving at a balanced (fibre) beam splitter (see inset of Fig.~\ref{fig:fig3}(c)) is~\cite{Branczyk2017}
\begin{equation}
    \label{eq:intra-pair coincidence probability}
    p(\tau)=\frac{1}{2} - \frac{1}{2} \int \int f^{\ast}(\omega_{\text{s}}, \omega_{\text{i}}) f(\omega_{\text{i}}, \omega_{\text{s}}) e^{i (\omega_{\text{i}} - \omega_{\text{s}})\tau} \mathrm{d}\omega_{\text{s}} \mathrm{d}\omega_{\text{i}}.
\end{equation}

To simulate the interference pattern in Fig.~\ref{fig:fig3}(c), the double integration is performed numerically.
To do so, we considered the measured $\sqrt{\text{JSI}}$ as shown in Fig.~\ref{fig:fig3}(b) to obtain a discrete (normalised) joint spectral amplitude $f_{\text{d}}(\omega_{\text{s}},\omega_{\text{i}}) \propto \exp \left(i\theta(\omega_{\text{s}},\omega_{\text{i}})\right)\sqrt{\text{JSI}(\omega_{\text{s}},\omega_{\text{i}})}$, where $\theta(\omega_{\text{s}},\omega_{\text{i}})$ applies the expected sign: $\theta(\omega_{\text{s}},\omega_{\text{i}})=\pi$ above the diagonal and $\theta(\omega_{\text{s}},\omega_{\text{i}})=0$ below the diagonal.

To obtain a fit function, we analytically evaluated Eq.~\eqref{eq:intra-pair coincidence probability} using Eqs.~\eqref{eq:PEF}-\eqref{eq:PMF}.
The double integration over the signal and idler frequencies results in a coincidence probability 
\begin{equation}
    p(\tau) = \frac{1}{2} +\frac{1}{2} \exp \left(-\frac{\sigma ^2 \tau ^2}{4}\right) \frac{\cos (\delta  \tau )-\eta}{1-\eta},
\end{equation}
with $\eta=\exp \left(-\delta ^2 /\sigma ^2 \right)$ representing the overlap between the two Gaussian peaks.

\begin{figure}
    \centering
    \includegraphics{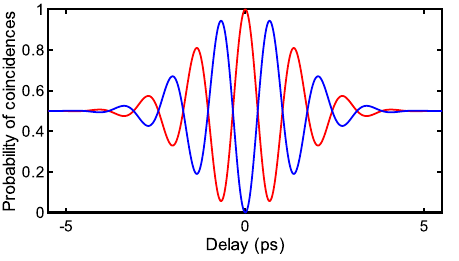}
    \caption{Analytical solutions of intra-pair HOM interference for PMFs where antinodes have a relative phase shift of $\pi$ (red line) and zero (blue line).}
    \label{fig:phase comparison}
\end{figure}

For well separated frequency bins ($\eta\ll1$, $\sim 10^{-7}$ in our case), this can be simplified to
\begin{equation}
\label{eq:HOM pi-phase shift}
    p(\tau) \approx \frac{1}{2} + \frac{1}{2} \exp \left(-\frac{\sigma ^2 \tau ^2}{4} \right)  \cos (\delta  \tau ),
\end{equation}
which generates a beating pattern with an oscillating term at the frequency-bin separation $\delta$, modulated by a Gaussian envelope.
This coincidence probability is used as fitting model for the intra-pair interference patterns with the bin width $\sigma$, their separation $\delta$, and the visibility (multiplicative factor in the range [0, 1] applied to the second term on the right-hand side) as free parameters.  

In the event of a relative zero phase shift being present between the Gaussian peaks of the phase-matching function, that is,
\begin{multline}
    \phi^{\prime}(\omega_{\text{s}}, \omega_{\text{i}}) = \exp \left(-\frac{(\omega_{\text{s}} - \omega_{\text{i}} + \delta )^2}{2 \sigma^2} \right) \\
    + \exp \left( -\frac{(\omega_{\text{s}} - \omega_{\text{i}} - \delta )^2}{2 \sigma^2} \right),
\end{multline}
the same calculations would result in a coincidence probability of
\begin{equation}
\label{eq:HOM 0-phase shift}
    p^{\prime}(\tau) \approx \frac{1}{2} - \frac{1}{2} \exp \left( -\frac{\sigma ^2 \tau ^2}{4} \right)  \cos (\delta  \tau ).
\end{equation}
Fig.~\ref{fig:phase comparison} shows a comparison between the interference patterns of Eqs.~\eqref{eq:HOM pi-phase shift} and \eqref{eq:HOM 0-phase shift}.
While a $\pi$-phase shift produces an anti-bunching peak at zero delay, a phase shift of zero results in a bunching dip at zero time delay.
This change in the interference behaviour is linked to the state's symmetry: a relative $\pi$ phase shift produces an antisymmetric state in the frequency DOF, whereas a zero phase shift results in a symmetric state.
In both cases, anti-bunching peaks can be observed at $\tau=\qty{0}{ps}$ or in its proximity, underlying the presence of frequency-bin entanglement.

\subsection{\label{appendix:inter pair fit model}Inter-pair interference}

From two independent, identical SPDC sources, each with joint spectrum $f(\omega_{\text{s}}, \omega_{\text{i}})$, heralding on the idler of each pair (see inset of Fig.~\ref{fig:fig3}(d)) projects the signal into a mixed spectral state~\cite{Branczyk2017}.
The reduced density matrix of each heralded signal photon is obtained by tracing over the idler:
\begin{align}
    \rho_{\text{s}}(\omega_{\text{s}}, \omega^{\prime}_{\text{s}}) = \int f(\omega_{\text{s}}, \omega_{\text{i}})f^{\ast}(\omega^{\prime}_{\text{s}}, \omega_{\text{i}}) \mathrm{d}\omega_{\text{i}}.
\end{align}
The heralded coincidence probability as a function of the relative time delay $\tau$ between the two photons arriving at a balanced (fibre) beam splitter becomes
\begin{align}
    \label{eq:inter-pair coincidence probability}
    p_{\text{H}}(\tau)=\frac{1}{2} - \frac{1}{2} \int \int \rho_{\text{s}}(\omega_{\text{s}}, \omega^{\prime}_{\text{s}}) \rho_{\text{s}}(\omega^{\prime}_{\text{s}}, \omega_{\text{s}}) e^{i (\omega^{\prime}_{\text{s}} - \omega_{\text{s}})\tau} \mathrm{d}\omega_{\text{s}} \mathrm{d}\omega^{\prime}_{\text{s}}.
\end{align}   

To simulate the interference pattern in Fig.~\ref{fig:fig3}(d), we numerically integrated the heralded coincidence probability, as discussed in the previous section.
To obtain a fit function, we performed the integration using again Eqs.~\eqref{eq:PEF}-\eqref{eq:PMF}.
By simplifying the results assuming $\eta \approx 0$ as before, the heralded coincidence probability becomes
\begin{align}
    p_{\text{H}}(\tau) \approx \frac{1}{2} - \frac{1}{16} \exp \left( -\frac{\sigma ^2 \tau ^2}{4} \right) \biggl(3 +  \cos (2 \delta  \tau ) \biggr),
\end{align}
where the oscillation frequency is twice that of the intra-pair interference in Eq.~\eqref{eq:HOM pi-phase shift}, due to the spectral intensity of the resulting heralded photons, which have peaks separated by $2\delta$.
This heralded coincidence probability is employed as fitting model for the inter-pair interference pattern of Fig.~\ref{fig:fig3}(d) with the bin width $\sigma$, their separation $\delta$, and the visibility (multiplicative factor in the range [0, 1] applied to the terms inside the brackets on the right-hand side) as free parameters.

\begin{figure}
    \centering
    \includegraphics[width=\linewidth]{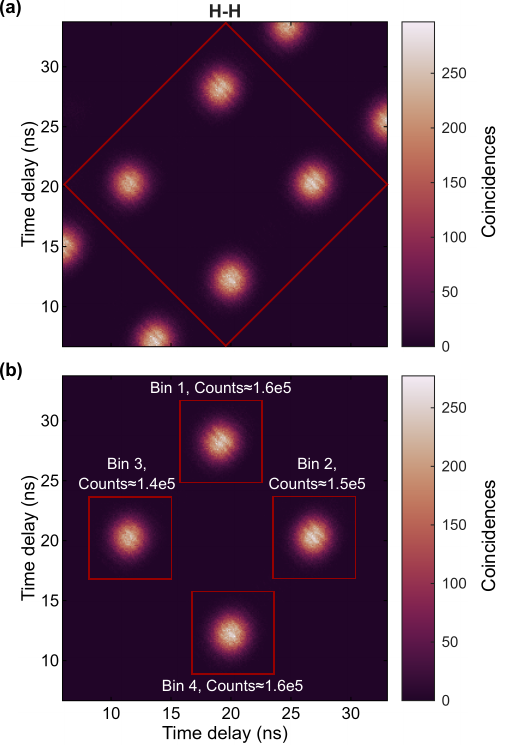}
    \caption{Polarisation-resolved TOFS analysis.
    \textbf{(a)} TOFS measurement for the example H-H polarisation projection.
    The red square indicates the region used to isolate the four-bin pattern from coincidence events associated with previous and subsequent trigger signals from the pulsed laser.
    Elements outside this area are set to zero.
    \textbf{(b)} The centres of the four frequency bins are identified and the total number of counts within a region of \qty{\pm 3.5}{ns} (red squares) is associated with the respective bin.
    }
    \label{fig:appendix_tomography}
\end{figure}

\section{\label{appendix:tomography}State Tomography}

The two-photon polarisation state tomography is based on 36 projection measurements (6 settings per photon: H, V, D, A, R, L) using a one-detector-per-qubit scheme, where only the transmitted outputs at the PBSs are considered~\cite{Altepeter2005}.
For each projection, the coincidence-time histogram is recorded, and the four-bin pattern is extracted as reported in Fig.~\ref{fig:appendix_tomography}(a).
This is used to evaluate the spectral purity (Schmidt number) for each polarisation projection presented in Fig.~\ref{fig:fig5}(e).
State tomography is then performed on the individual bins by integrating the total number of counts over a \qty{\pm 3.5}{ns} region centred around each frequency bin, as shown in Fig.~\ref{fig:appendix_tomography}(b).
The isolated areas are large enough to fully include the bin without overlapping with other bins.
Maximum likelihood estimation reconstructs the density matrix for each bin.
Uncertainties are estimated via Monte Carlo simulation assuming Poissonian counting statistics for the recorded coincidence-time histograms.

\section{\label{appendix:polarisation resolved HOM}Polarisation-resolved HOM}

We assume that the hyperentangled state can be written as the tensor product between the polarisation state $\ket{\psi_{\text{p}}}$ and frequency state $\ket{\psi_{\text{f}}}$:
\begin{align}
\label{eq:total state}
    \ket{\psi} = \ket{\psi_{\text{p}}} \otimes \ket{\psi_{\text{f}}},
\end{align}
where
\begin{align}
    \ket{\psi_{\text{p}}} = \frac{1}{\sqrt{2}} \left( \ket{\text{HV}}_{\text{s,i}} + e^{i\varphi}\ket{\text{VH}}_{\text{s,i}}\right)
\end{align}
and
\begin{multline}
\label{eq:frequency state}
    \ket{\psi_{\text{f}}} = \frac{1}{2} \biggl[ \left(\ket{\omega_2,\omega_3}_{\text{s,i}} -\ket{\omega_3,\omega_2}_{\text{s,i}} \right) \\
    + \left( \ket{\omega_1,\omega_2}_{\text{s,i}} -\ket{\omega_2,\omega_1}_{\text{s,i}} \right)  \biggr],
\end{multline}
with $\omega_1 > \omega_2 > \omega_3$ as the central frequencies of each frequency bin in the marginal distributions (see Fig.~\ref{fig:fig1}(c)).
The polarisation Bell states $\ket{\psi^{+}_{\text{p}}}$ (triplet state) and $\ket{\psi^{-}_{\text{p}}}$ (singlet state) correspond to phase factors $\varphi=0$ and $\varphi=\pi$, respectively, and have opposite symmetries with respect to permutations of the subsystems (exchanging $\text{s} \rightleftharpoons \text{i}$): $\ket{\psi^{+}_{\text{p}}}$ is symmetric while $\ket{\psi^{-}_{\text{p}}}$ is antisymmetric.
The frequency state $\ket{\psi_{\text{f}}}$, written in Eq.~\eqref{eq:frequency state} as a sum of two antisymmetric Bell states, is also antisymmetric under exchange.
Consequently, the total hyperentangled state $\ket{\psi}$ is antisymmetric when $\ket{\psi_{\text{p}}}=\ket{\psi^{+}_{\text{p}}}$ and symmetric when $\ket{\psi_{\text{p}}}=\ket{\psi^{-}_{\text{p}}}$.

This symmetry can be probed through two-photon interference at a balanced beam splitter with polarisation-resolved detection, as shown in Fig.~\ref{fig:fig6}(a).
Since indistinguishable bosons must obey Bose-Einstein statistics, their total wavefunction must be symmetric under particle exchange.
Interference at the beam splitter converts exchange symmetry into spatial correlations: symmetric states lead to photon bunching, while antisymmetric states lead to antibunching~\cite{Fedrizzi2009}.

For our hyperentangled state, this implies that photons should bunch when $\ket{\psi_{\text{p}}}=\ket{\psi^{-}_{\text{p}}}$ and antibunch when $\ket{\psi_{\text{p}}}=\ket{\psi^{+}_{\text{p}}}$.
This behaviour is the opposite of that observed with ordinarily polarisation-entangled photons where the triplet state would bunch, but the additional antisymmetric structure of the frequency-bin entanglement reverses this expectation. Detecting coincidences between different output ports while performing polarisation projections in the horizontal/vertical (H/V) and diagonal/antidiagonal (D/A) bases directly verifies this prediction.

To further support this symmetry-based prediction, we directly model the HOM interference at the beam splitter of signal and idler photons of the state in Eq.~\eqref{eq:total state}.
In the H/V polarisation basis, the input state before the beam splitter can be written as 
\begin{widetext}
\begin{align}
\label{eq:input state}
    \ket{\psi}_{\text{in}} = \frac{1}{2\sqrt{2}} \left[
    a^{\dagger}_{\text{H},2} b^{\dagger}_{\text{V},3} - a^{\dagger}_{\text{H},3} b^{\dagger}_{\text{V},2} + a^{\dagger}_{\text{H},1} b^{\dagger}_{\text{V},2} - a^{\dagger}_{\text{H},2} b^{\dagger}_{\text{V},1}
    + e^{i\varphi} \left(
    a^{\dagger}_{\text{V},2} b^{\dagger}_{\text{H},3} - a^{\dagger}_{\text{V},3} b^{\dagger}_{\text{H},2} + a^{\dagger}_{\text{V},1} b^{\dagger}_{\text{H},2} - a^{\dagger}_{\text{V},2} b^{\dagger}_{\text{H},1}
    \right)
    \right] \ket{0}_{\text{s,i}},
\end{align}
\end{widetext}
with $a^{\dagger}_{j,k}$ and $b^{\dagger}_{j,k}$ the bosonic creation operators generating a photon in spatial modes $a$ and $b$ (the two input ports of the beam splitter), respectively, with polarisation $j=\text{H}, \, \text{V}$ and frequency $\omega_k$, where $k=1, \, 2, \,3 $.
In these calculations, we consider perfect temporal overlap of the two photons at the beam splitter, corresponding to the zero-delay condition in the experiment. 
Assuming a balanced splitting, the beam splitter unitary $U_{\text{BS}}$ transforms the creation operators to
\begin{align}
    \label{eq:BS1}
    a^{\dagger}_{j,k} & \longrightarrow \frac{1}{\sqrt{2}}\left( a^{\dagger}_{j,k} + b^{\dagger}_{j,k}  \right)\\
    \label{eq:BS2}
    b^{\dagger}_{j,k} & \longrightarrow \frac{1}{\sqrt{2}}\left( a^{\dagger}_{j,k} - b^{\dagger}_{j,k}  \right). 
\end{align}
By applying Eqs.~\eqref{eq:BS1}-\eqref{eq:BS2} to Eq.~\eqref{eq:input state} and reordering the terms, the state at the output of the beam splitter is
\begin{widetext}
\begin{multline}
\label{eq:output state H/V}
\ket{\psi}_{\text{out}}  = \frac{1}{4\sqrt{2}} \biggl[ (1-e^{i\varphi}) \left(
    a^{\dagger}_{\text{H},2} a^{\dagger}_{\text{V},3} - a^{\dagger}_{\text{H},2} a^{\dagger}_{\text{V},1} - a^{\dagger}_{\text{H},3} a^{\dagger}_{\text{V},2} + a^{\dagger}_{\text{H},1} a^{\dagger}_{\text{V},2} -
    b^{\dagger}_{\text{H},2} b^{\dagger}_{\text{V},3} + b^{\dagger}_{\text{H},2} b^{\dagger}_{\text{V},1} + b^{\dagger}_{\text{H},3} b^{\dagger}_{\text{V},2} - b^{\dagger}_{\text{H},1} b^{\dagger}_{\text{V},2}
    \right) \\
     -(1+e^{i\varphi}) \left(
    a^{\dagger}_{\text{H},2} b^{\dagger}_{\text{V},3} - a^{\dagger}_{\text{V},3} b^{\dagger}_{\text{H},2} - a^{\dagger}_{\text{H},2} b^{\dagger}_{\text{V},1} + a^{\dagger}_{\text{V},1} b^{\dagger}_{\text{H},2} - a^{\dagger}_{\text{H},3} b^{\dagger}_{\text{V},2} + a^{\dagger}_{\text{V},2} b^{\dagger}_{\text{H},3} + a^{\dagger}_{\text{H},1} b^{\dagger}_{\text{V},2} - a^{\dagger}_{\text{V},2} b^{\dagger}_{\text{H},1}
    \right)
    \biggr] \ket{0}_{\text{s,i}}.
\end{multline}
\end{widetext}

This state corresponds to photons bunching or antibunching depending on the phase factor $\varphi$.
When $\varphi=0$ (triplet state $\ket{\psi^+_{\text{p}}}$), the prefactor $(1-e^{i\varphi})$ vanishes, leaving only terms of the form $a^{\dagger}_{j,k} b^{\dagger}_{j',k'}$, corresponding to antibunching.
Contrarily, when $\varphi=\pi$ (singlet state $\ket{\psi^-_{\text{p}}}$), the prefactor $(1+e^{i\varphi})$ vanishes, leaving only terms of the form $a^{\dagger}_{j,k} a^{\dagger}_{j',k'}$ or $b^{\dagger}_{j,k} b^{\dagger}_{j',k'}$, corresponding to bunching.
In both cases, the two output photons always have orthogonal polarisation states.
These two results are reflected in the experimental results shown in Fig.~\ref{fig:fig6}(a)-(b) at the zero-delay condition.

However, the H/V measurement alone cannot confirm the antisymmetric character of the frequency-bin state.
To provide direct evidence that antibunching corresponds to the triplet state and bunching to the singlet state, thereby confirming the essential role of the antisymmetric frequency-bin entanglement, we analyse the output state in the D/A basis, with orthonormal state vectors $\ket{\text{D}}=(\ket{\text{H}}+\ket{\text{V}})/\sqrt{2}$ and $\ket{\text{A}}=(\ket{\text{H}}-\ket{\text{V}})/\sqrt{2}$.
By expressing the output state in Eq.~\eqref{eq:output state H/V} in this basis, we find
\begin{widetext}
\begin{multline}
\label{eq:output state D/A}
\ket{\psi}_{\text{out}}  = \frac{1}{4\sqrt{2}} \biggl[ (1-e^{i\varphi}) \left(
    a^{\dagger}_{\text{A},2} a^{\dagger}_{\text{D},3} - a^{\dagger}_{\text{A},2} a^{\dagger}_{\text{D},1} - a^{\dagger}_{\text{A},3} a^{\dagger}_{\text{D},2} + a^{\dagger}_{\text{A},1} a^{\dagger}_{\text{D},2} -
    b^{\dagger}_{\text{A},2} b^{\dagger}_{\text{D},3} + b^{\dagger}_{\text{A},2} b^{\dagger}_{\text{D},1} + b^{\dagger}_{\text{A},3} b^{\dagger}_{\text{D},2} - b^{\dagger}_{\text{A},1} b^{\dagger}_{\text{D},2}
    \right) \\
     -(1+e^{i\varphi}) \left(
    a^{\dagger}_{\text{A},2} b^{\dagger}_{\text{A},1} - a^{\dagger}_{\text{A},1} b^{\dagger}_{\text{A},2} - a^{\dagger}_{\text{A},2} b^{\dagger}_{\text{A},3} + a^{\dagger}_{\text{A},3} b^{\dagger}_{\text{A},2} - a^{\dagger}_{\text{D},2} b^{\dagger}_{\text{D},1} + a^{\dagger}_{\text{D},1} b^{\dagger}_{\text{D},2} + a^{\dagger}_{\text{D},2} b^{\dagger}_{\text{D},3} - a^{\dagger}_{\text{D},3} b^{\dagger}_{\text{D},2}
    \right)
    \biggr] \ket{0}_{\text{s,i}}.
\end{multline}
\end{widetext}
The change of basis does not affect the antibunching and bunching behaviours, which still correspond to phase factors $\varphi=0$ and $\varphi=\pi$, respectively.
However, the projection in the D/A basis makes the triplet and singlet states distinguishable.
Crucially, while the singlet state ($\varphi=\pi$) still produces photons with orthogonal polarisations as in the H/V basis, the triplet state ($\varphi=0$) now produces photons with identical polarisations.
This qualitative difference provides an unambiguous experimental signature, which is confirmed by the experimental results shown in Fig.~\ref{fig:fig6}(c)-(d): at zero delay, the triplet state $\ket{\psi^+_{\text{p}}}$ shows antibunching with coincidences only in same-polarisation outputs (TT and RR), while the singlet state $\ket{\psi^-_{\text{p}}}$ exhibits coincidences in cross-polarisation outputs (TR and RT).

The observation of antibunching for $\ket{\psi^+_{\text{p}}}$ and bunching for $\ket{\psi^-_{\text{p}}}$, opposite to what would occur for polarisation entanglement alone, combined with the basis-dependent polarisation correlations, constitutes direct evidence that the frequency-bin DOF carries an antisymmetric entangled state, confirming the hyperentanglement generated by our source.

\vspace{0.5cm}


\end{document}